# Design of Reversible Random Access Memory

Md. Selim Al Mamun
Department of Computer Science and Engineering,
Jatiya Kabi Kazi Nazrul Islam University,
Bangladesh.

Syed Monowar Hossain
Department of Computer Science and Engineering,
University of Dhaka
Bangladesh.

## ABSTRACT

Reversible logic has become immensely popular research area and its applications have spread in various technologies for their low power consumption. In this paper we proposed an efficient design of random access memory using reversible logic. In the way of designing the reversible random access memory we proposed a reversible decoder and a write enable reversible master slave D flip-flop. All the reversible designs are superior in terms of quantum cost, delay and garbage outputs compared to the designs existing in literature.

## 1. INTRODUCTION

In recent year reversible computing has emerged as a promising technology. The primary reason for this is the increasing demands for low power devices. R. Landauer [1] proved that losing information causes loss of energy. Information is lost when an input cannot be recovered from its output. He showed that each bit of information loss generates kTln2 joules of heat energy; where k is Boltzmann's constant and T is the absolute temperature at which computation is performed. C. H. Bennett [2] showed that energy dissipation problem can be avoided if circuits are built using reversible logic gates.

In reversible logic there is a one to one mapping between its input and output. As a result no information bit is lost and no loss of energy [3]. Random Access Memory (RAM) uses bistable sequential circuitry to store a single bit. Although many researchers are working on reversible memory elements, little work has been done in this area. In current literature the number of reversible gates is used as a major metric of cost optimization [4]. Dmitri Maslov and Michael Miller [5] showed that number of gates is not a good metric of optimization as reversible gates are of different types and have different quantum costs. In this paper, we proposed a new design of RRAM that is efficient in terms of quantum cost, delay and the number of garbage outputs.

The rest of the paper is organized as follows: Section 2 presents some basic definitions related to reversible logic. Section 3 describes some popular reversible logic gates and their quantum representation. Section 4 describes our proposed modification on Frekdin gate (FRG). Section 5 describes the logic synthesis of RRAM and compares our proposed design with other researchers. Finally this paper is concluded with Section 6.

## 2. BASIC DEFINITIONS

In this section, some basic definitions related to reversible logic are presented. We formally define reversible gate, garbage output, delay and quantum cost in reversible circuits.

### 2.1. Reversible Gate

A Reversible Gate is a k-input and k-output (denoted by k*k) circuit that produces a unique output pattern for each possible input pattern [6]. If the input vector of the reversible gate is defined as $Iv$ where $Iv = (I_{1,j}, I_{2,j}, I_{3,j}, ...., I_{k-1,j}, I_{k,j})$ and the output vector as $Ov$ where $Ov = (O_{1,j}, O_{2,j}, O_{3,j}, ..., O_{k-1,j}, O_{k,j})$, then according to the definition, for each particular vector $j$, $Iv \leftrightarrow Ov$.

### 2.2. Garbage Output

Outputs that are not primary outputs or outputs that are not used as input to other gates to produce primary outputs are garbage. Unwanted or unused outputs which are needed to maintain reversibility of a reversible gate (or circuit) are known as Garbage Outputs. The garbage output of Feynman gate [7] is shown Figure 1 with *.

### 2.3. Delay

The delay of a logic circuit is the maximum number of gates in a path from any input line to any output line. The definition is based on two assumptions: (i) Every gate computation takes one unit of time and (ii) All inputs to the circuit are available before the computation. In this paper, we used the logical depth as measure of the delay proposed by Mohammadi and Eshghi [8]. The delay of each 1x1 gate and 2x2 reversible gate is taken as unit delay 1. Any 3x3 reversible gate can be designed from 1x1 reversible gates and 2x2 reversible gates, such as CNOT gate, Controlled-V and Controlled-$V^+$ gates (V is a square-root-of NOT gate and $V^+$ is its hermitian). Thus, the delay of a 3x3 reversible gate can be computed by calculating its logical depth when it is designed from smaller 1x1 and 2x2 reversible gates.

### 2.4. Quantum Cost

The quantum cost of a reversible gate is defined as the number of 1x1 and 2x2 reversible gates or quantum gates needed to realize the design. The quantum costs of all reversible 1x1 and 2x2 gates are taken as unity [9]. Since every reversible gate is a combination of 1x1 or 2x2 quantum gate, the quantum cost of any reversible gate can be calculated by counting the numbers of NOT, Controlled-V, Controlled-$V^+$ and CNOT gates used in the design.

## 3. QUANTUM ANALYSIS OF POPULAR REVERSIBLE GATES

Every reversible gate can be realized by the quantum gates. Thus the cost of reversible circuit can be measured in terms of quantum cost. Reducing the quantum cost of a reversible circuit is always a challenging one and works are still going on in this area. This section describes some popular reversible gates and presents quantum equivalent diagram of each of the reversible gate.

### 3.1 Feynman Gate

The input vector $I_v$ and output vector $O_v$ of 2*2 Feynman gate is defined as $I_v = (A, B)$ and $O_v = (P = A, Q = A \oplus B)$. The quantum cost of Feynman gate is 1. The block diagram and equivalent quantum representation of 2*2 Feynman gate are shown in Fig. 1.

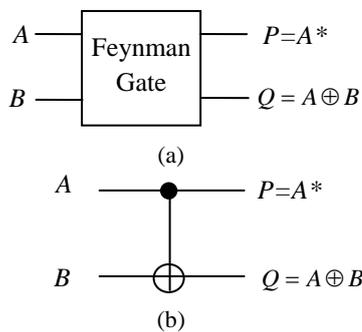

**Fig. 1. (a) Block diagram of 2x2 Feynman gate and (b) Equivalent quantum representation**

### 3.2 Double Feynman Gate

The input vector $I_v$ and output vector $O_v$ of 3*3 Double Feynman gate (DFG) is defined as $I_v = (A, B, C)$ and $O_v = (P = A, Q = A \oplus B, R = A \oplus C)$. The quantum cost of Double Feynman gate is 2 [10]. The block diagram and equivalent quantum representation of 3*3 Double Feynman gate are shown in Fig. 2.

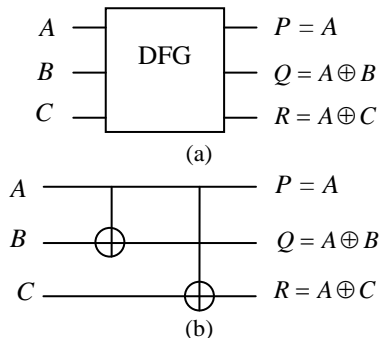

**Fig. 2. (a) Block diagram of 3x3 Double Feynman gate and (b) Equivalent quantum representation.**

### 3.3 Toffoli Gate

The input vector $I_v$ and output vector $O_v$ of 3*3 Toffoli gate (TG) [11] is defined as $I_v = (A, B, C)$ and $O_v = (P = A, Q = B, R = AB \oplus C)$. The quantum cost of Toffoli gate is 5. The block diagram and equivalent quantum representation of 3*3 Toffoli gate are shown in Fig. 3.

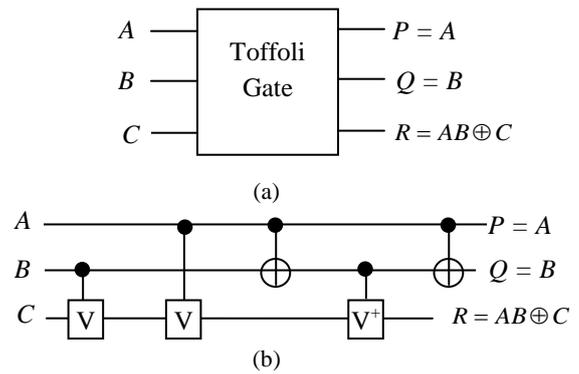

**Fig.3. (a) Block diagram of 3*3 Toffoli gate and (b) Equivalent quantum representation.**

### 3.4 Frekdin Gate

The input vector $I_v$ and output vector $O_v$ for 3*3 Fredkin gate (FRG) [12] is defined as follows: $I_v = (A, B, C)$ and $O_v = (P=A, Q = \overline{A}B \oplus AC, R = \overline{A}C \oplus AB)$. The quantum cost of Frekdin gate is 5. The block diagram and equivalent quantum representation of 3*3 Fredkin gate are shown in Fig. 4.

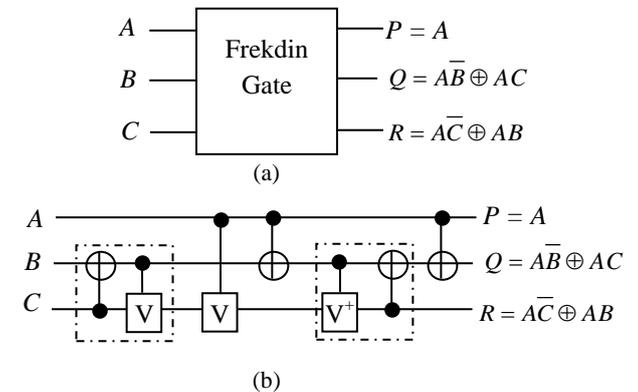

**Fig.4. (a) Block diagram of 3*3 Frekdin gate and (b) Equivalent quantum representation.**

### 3.5 Peres Gate

The input vector $I_v$ and output vector $O_v$ of 3*3 Peres gate (PG)[13] is defined as follows: $I_v = (A, B, C)$ and $O_v = (P = A, Q = A \oplus B, R = AB \oplus C)$. The quantum cost of Peres gate is 4. The block diagram and equivalent quantum representation of 3*3 Peres gate are shown in Fig. 5.

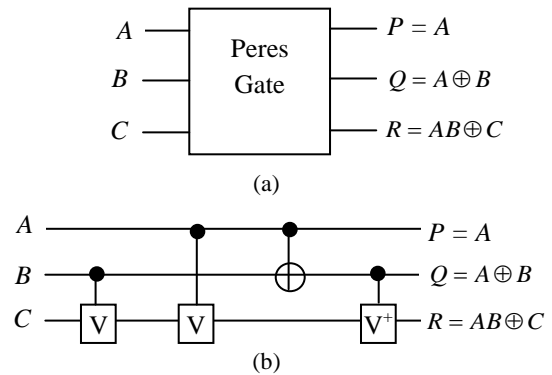

**Fig.5. (a) Block diagram of 3*3 Peres and (b) Equivalent quantum representation.**

## 4. PROPOSED MODIFICATION ON FREKDIN GATE

### 4.1 Modified FRG 1 gate

The input vector, $I_v$ and output vector, $O_v$ for 3*3 modified Fredkin Gate (MFRG1) is defined as follows: $I_v = (A, B, C)$ and $O_v = (P=A, Q = \overline{A}B \oplus A\overline{C}, R = \overline{A}C \oplus AB)$. The quantum cost of MFRG1 gate is 4.

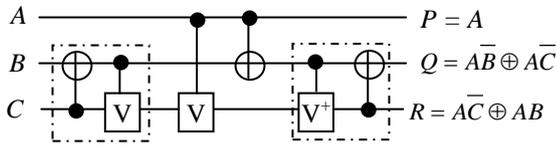

**Fig. 6. Quantum representation of MFRG1 gate**

### 4.2 Modified FRG 2 gate

The input vector $I_v$ and output vector $O_v$ of 3*3 modified Fredkin Gate (MFRG2) is defined as $I_v = (A, B, C)$ and $O_v = (P = \overline{A}, Q = \overline{A}B \oplus AC, R = \overline{A}C \oplus AB)$. The quantum cost of MFRG2 gate is 5.

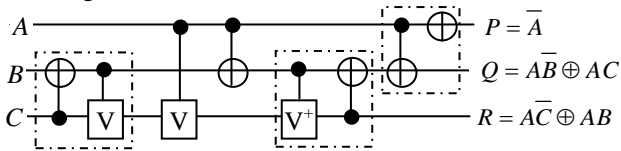

**Fig. 7. Quantum representation of MFRG2 gate**

## 5. DESIGN OF RANDOM ACCESS MEMORY

In this section we first presented proposed design for all the components of RRAM. Then we presented our proposed novel design of RRAM that is optimized in terms of quantum cost, delay and garbage outputs.

### 5.1. Proposed Reversible $n$ to $2^n$ Decoder

A single Feynman gate can be used to design the basic *1 to $2^1$* decoder. Using this decoder we can systematically add $2^n$-1 number of MRFG1 gates to the design to achieve *n to $2^n$* decoder. The design of *1 to $2^1$* decoder is shown in Figure. 8.

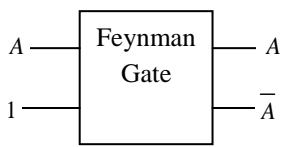

**Fig. 8. Proposed 1 to $2^1$ decoder**

The design of decoder has 1 quantum cost, 1 delay and no garbage output.

Our proposed *2 to $2^4$* decoder using MRFG1 gates are shown in Figure 9.

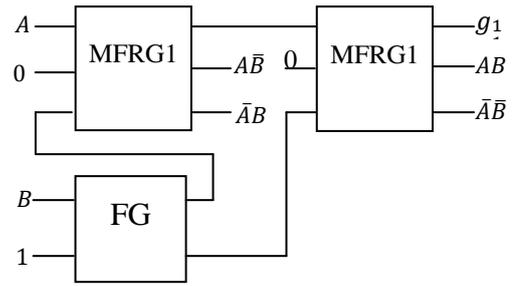

**Fig. 9. Proposed *2 to $2^2$* decoder**

The proposed *2 to $2^2$* decoder has quantum cost 9, delay 9 and bare minimum of 1 garbage bit. The proposed design of *2 to $2^2$* decoder achieves improvement ratios of 18%, 18% and 50% in terms of quantum cost, delay and garbage outputs compared to the design presented in N.M.Nayeem et al. [14]. The improvement ratios compared to the design M. Morrison et al. presented in [19] are 10% and 10% in terms of quantum cost and delay. The comparison of proposed *2 to $2^2$* decoder with the existing ones shown in table I.

**Table I. Comparison of different types of 2 to $2^2$ decoders.**

| 2 to $2^2$ decoder design | Cost Comparisons | | |
|---|---|---|---|
| | Quantum Cost | Delay | Garbage Outputs |
| **Proposed** | 9 | 9 | 1 |
| Existing[14] | 11 | 11 | 2 |
| Existing[19] | 10 | 10 | - |
| Improvement(%) w.r.t. [14] | 18 | 18 | 50 |
| Improvement(%) w.r.t. [19] | 10 | 10 | - |

**Theorem 1:** To construct *n to $2^n$* decoder, if $g$ is the total number of gates required to design the decoder producing $b$ number of garbage outputs then $g \geq 2^n-1$ and $b \geq n-1$.

**Proof:** For *1 to $2^1$* decoder only one Feynman gate needed that doesn't produce any garbage bit. So number of gate = 1 and garbage output = 0.

Now for n>1, *n to $2^n$* decoder design requires that each of the output of the *(n-1) to $2^{(n-1)}$* decoders together with (n-1) selection bits are employed in separate MRFG1 gate to produce selections for n to $2^n$ decoder. In that case overall number of gates becomes $2^n-1$, because for n=1 we were required only 1 gate. This design has n-1 garbage bits as the 1 to $2^1$ decoder produces zero garbage.

**Theorem 2:** The quantum cost of an *n to $2^n$* decoder is $Q_c \geq 4.2^n - 7$.

**Proof:** From theorem 1, at least $2^n-1$ gates are required to design n to $2^n$ decoder. For n = 1 only one 2*2 Feynman gate is required which has quantum cost 1. Then $2^n-2$ MRFG1 gates are required and each MRF1 gate's quantum cost is 4. So total quantum cost $Q_c$ is = $(2^n-2)4 +1 = 4.2^n-7$.

### 5.2. Proposed Single bit Memory Cell

The heart of our proposed memory block is D flip-flop. The characteristic equation of gated D flip-flop is $Q^+ = CLK.D + \overline{CLK}.Q$. The D flip-flop can be realized by one MFRG2 gate and one FG. It can be mapped with MFRG2 by giving *CLK*, D and Q respectively in $1^{st}$, $2^{nd}$ and $3^{rd}$ inputs of MFRG2 gate. The Figure 10 shows our proposed D flip-flop with $Q$ and $\bar{Q}$ outputs.

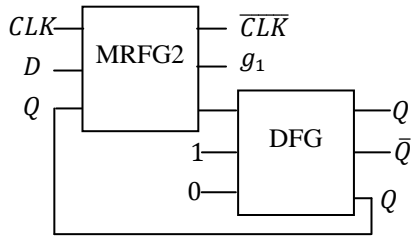

**Fig. 10: Proposed design of D flip-flop with $Q$ and $\bar{Q}$ outputs.**

The proposed D flip-flop with $Q$ and $\bar{Q}$ outputs has quantum cost 7, delay 7 and has the bare minimum of 1 garbage bit. The proposed design of gated D flip-flop achieves improvement ratios of 50% in terms of garbage outputs compared to the design presented in Thapliyal et al. 2010[15] and L. Jamal et al. 2012[16]. The comparisons of our D flip-flop (with $Q$ and $\bar{Q}$ outputs) design with existing designs in literature are summarized in Table II.

**Table II. Comparison of different types of D flip-flops with $Q$ and $\bar{Q}$ outputs.**

| D flip-flop design | Cost Comparisons | | |
|---|---|---|---|
| | Quantum Cost | Delay | Garbage Outputs |
| **Proposed** | 7 | 7 | 1 |
| Existing[15] | 7 | 7 | 2 |
| Existing[16] | 7 | 7 | 2 |
| Improvement(%) w.r.t. [15] | 0 | 0 | 50 |
| Improvement(%) w.r.t. [16] | 0 | 0 | 50 |

We need Write Enable Master Slave D FFs to design a Reversible Random Access Memory (RRAM). Our proposed write enable master slave flip-flop is shown in Figure 11.

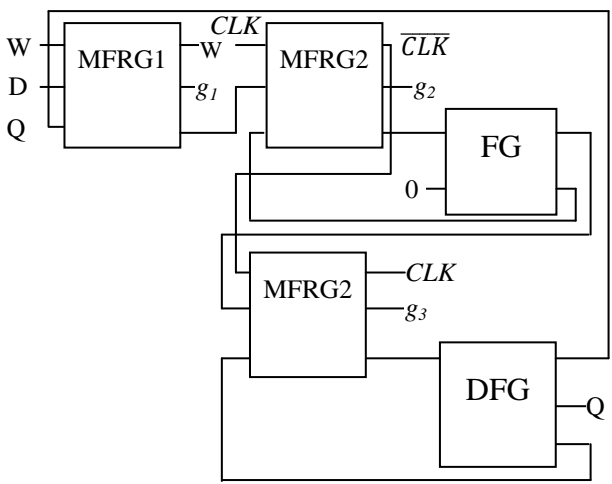

**Fig. 11: Proposed write enable master slave D flip-flop.**

As data are both read from and written into RAM, each Flip-Flop should work on two modes- read and write. A MFRG1 gate is used to multiplex between flip-flop's $D$ input and stored bit $Q$ in the flip-flop. When 'write' is high, input $D$ of the MRFG1 gate is carried to the first $D$ flip-flop and if 'write' is low, output of the D flip-flop is fed back to the second input of the MRFG1 gate so that state of the flip-flop remains same.

The proposed write enable master slave D flip-flop has quantum cost 17, delay 17 and 3 garbage bits. The proposed design of single bit memory cell achieves improvement ratios of 19% and 11% in terms of quantum cost and delay compared to the design presented in M. Morrison [19]. The comparisons are summarized in table III.

**Table III. Comparisons of different types of Write Enable Master Slave D flip-flops with $Q$ and $\bar{Q}$ outputs.**

| Single bit Memory Cell | Cost Comparisons | | |
|---|---|---|---|
| | Quantum Cost | Delay | Garbage Outputs |
| **Proposed** | 17 | 17 | 3 |
| Existing[19] | 21 | 19 | - |
| Improvement in (%) | 19 | 11 | - |

## 5.3. Proposed Reversible Random Access Memory (RRAM)

A RAM is a two dimensional array of flip-flops. There are $2^n$ rows where each row contains $m$ flip-flops. Each time only one of the $2^n$ output lines of the decoder is active which selects one row of flip-flops of the RAM. Whether a read or a write operation is performed depends on the $W$ input. When $W$ is high, $m$ flip-flops of the selected row of the RAM are written with the inputs $D_1$ to $D_m$. When $W$ is low, $Q_1$ to $Q_m$ contains stored bits in the flip-flops of the selected row and simultaneously the flip-flops are refreshed with the stored bits. The proposed design of $2^n*m$ bit RRAM is shown in Figure.12.

**Theorem 3**
Let $g$ be the number of gates required to realize a $2^n * m$ Reversible RAM where $n$ be the number of bits and $m$ be the selection bits in the RRAM, then $g \geq 2^n *(6m+2) + m-1$.

**Proof**: A $2^n * m$ RRAM requires $n$ to $2^n$ decoder that consists of $(2^n-1)$ gates. $2^n$ Toffoli gates are required to perform AND operations in RRAM. $m*2^n$ D flip-flops are required inside the $m * 2^n$ RRAM whereas each D flip-flop requires 5 gates. $2^n * m$ Feynman gates are required to perform the copy operation. There are m number of $2^n$ bit Feynman gate at the blottom last row. If $g$ be the minimum number of gates to realize the RRAM, then $g \geq (2^n-1) + 2^n + 5*2^n*m + 2^n*m+m$
Hence $g \geq 2^n *(6m+2)+m -1$.

**Theorem 4**
Let $n$ be the number of bits, $m$ be the selection bits in the RRAM and b be the number of garbage outputs generated from the RRAM, then $b \geq m*(4.2^n -1)+n$.

**Proof:** A $2^n * m$ RRAM requires $n$ to $2^n$ decoder which produces (n-1) garbage bits. 1 garbage bit is generated from the $m^{th}$ Toffoli gate in the RRAM. Inside RRAM there are $2^n*m$ D flip-flops and each D flip-flop produces 3 garbage bits. The last row contains m number of $2^n$ bit Feynman gate and each of them produces $2^n-1$ garbage bits. If $b$ be the number of garbage outputs then $b \geq (n-1) + 1 + 3*2^n*m + m*(2^n-1)$.
Hence $b \geq m*(4.2^n -1) + n$.

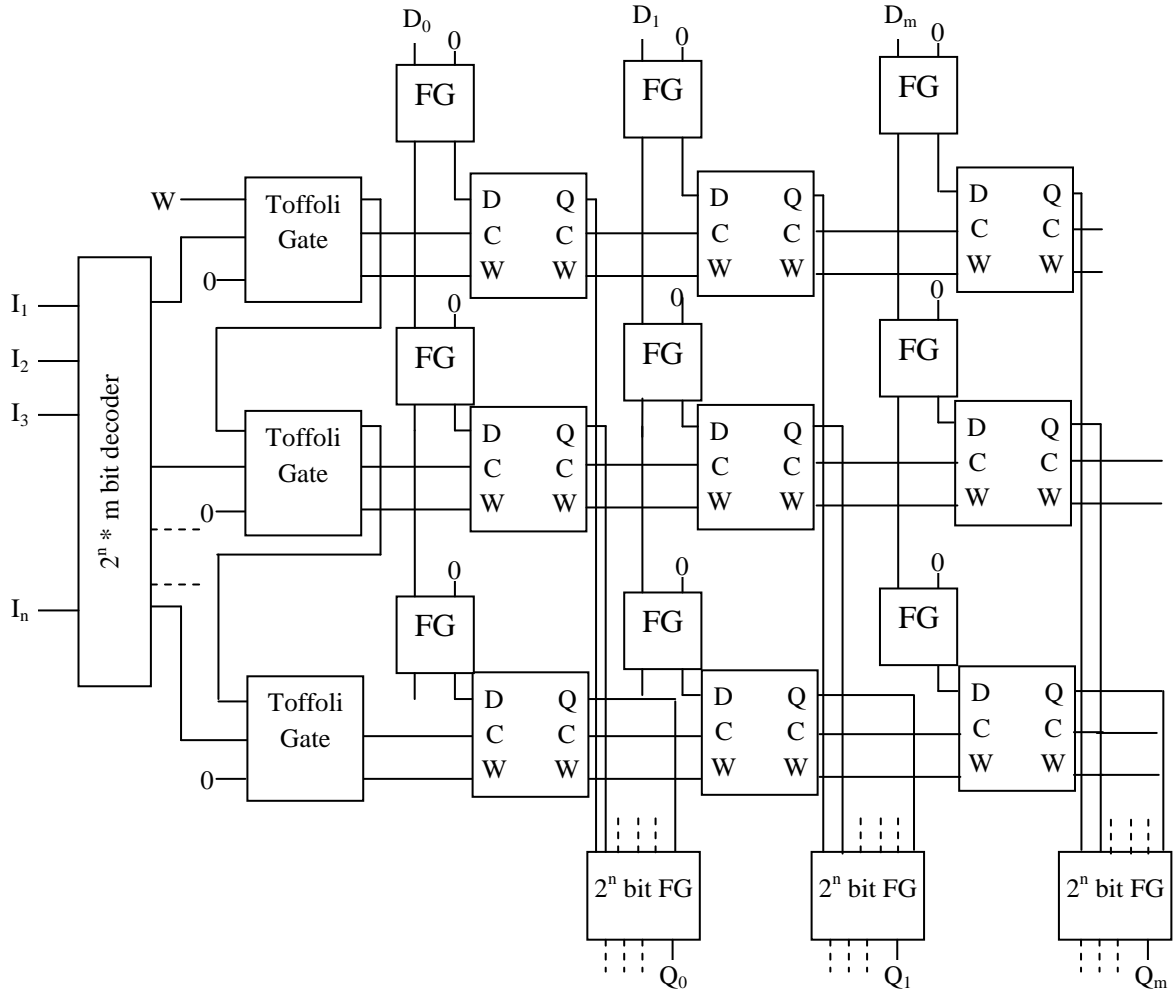

**Fig.12. Proposed design of $2^n * m$ bit RRAM**

**Theorem 5**
Let *n* be the number of bits, *m* be the selection bits in the RRAM and $Q_c$ be the quantum cost of the RRAM, then $Q_c \geq 2^n(19m+9)-7$.

**Proof:** A $2^n * m$ RRAM requires an *n to $2^n$* decoders that has quantum cost $4.2^n-7$. $2^n$ Toffoli gates are used to perform AND operations in RRAM where each of the gates has quantum cost 5, $m*2^n$ DFFs are required inside the $m * 2^n$ RRAM whereas each DFF has quantum cost 17. $2^n * m$ Feynman gates are required to perform the copy operation where each of them has quatum cost 1. There are m number of $2^n$ bit Feynman gate in last row where each of them has quantum cost $2^n$. If Qc be the quantum cost of RRAM then $Q_c \geq 4.2^n-7 +5.2^n+ m*2^n+17*m*2^n+ m*2^n$.
Hence $Q_c \geq 2^n(19m+9)-7$

## 6. CONCLUSION
Reversible Random Access Memory (RRAM) is going to take the place of existing main memory in the forthcoming quantum devices. In this paper we proposed optimized RRAM with the help of proposed MRFG1 and MRFG2 gates along with some basic reversible logic gates. Appropriate algorithms and theorems are presented here to clarify the proposed design and to establish its efficiency. We compare our design with existing ones in literature which claims our success in terms of quantum cost, number of garbage outputs and delay. We believe this optimization can contribute significantly in reversible logic community.

## 7. ACKNOWLEDGMENTS
The authors would like to thank the anonymous referees for their constructive feedback, which helped significantly improving technical quality of this paper.